\documentclass[9pt,conference]{IEEEtran}

\IEEEoverridecommandlockouts

\usepackage{amsmath}
\usepackage{amsthm}
\usepackage{amsfonts}
\usepackage{amssymb}
\usepackage{mathtools}
\usepackage{float}
\usepackage{algorithm,algorithmic}
\usepackage[pdftex]{graphicx}
\usepackage{amsmath}
\usepackage{color}
\usepackage{multirow}
\usepackage{graphicx}
\usepackage{bm}
\usepackage{balance}
\usepackage{comment}
\usepackage[normalem]{ulem} 

\usepackage[inline]{enumitem}

\usepackage{cite}

\usepackage[caption=false,font=footnotesize]{subfig}

\newcommand{\bsym}{\boldsymbol}

\DeclareMathOperator*{\minimize}{minimize }

\newtheorem{theorem}{Theorem}

\newtheorem{proposition}[theorem]{Proposition}

\begin{document}
\title{CLuP-Based Dual-Deconvolution in \\Automotive ISAC Scenarios}
\author{Jonathan Monsalve and Kumar Vijay Mishra
\thanks{J. M. is with the Universidad Industrial de Santander, Bucaramanga, Santander 680002 Colombia. Email: jonathan.monsalve@correo.uis.edu.co.}
\thanks{K. V. M. is with the United States DEVCOM Army Research Laboratory, Adelphi, MD 20783 USA. E-mail: kvm@ieee.org.}
\thanks{K. V. M. acknowledges support from the National Academies of Sciences, Engineering, and Medicine via the Army Research Laboratory Harry Diamond Distinguished Fellowship.}}

\maketitle

\begin{abstract}
Accurate target parameter estimation of range, velocity, and angle is essential for vehicle safety in advanced driver assistance systems (ADAS) and autonomous vehicles. To enable spectrum sharing, ADAS may employ integrated sensing and communications (ISAC). This paper examines a dual-deconvolution automotive ISAC scenario where the radar waveform is known but the propagation channel is not, while in the communications domain, the channel is known but the transmitted message is not. Conventional maximum likelihood (ML) estimation for automotive target parameters is computationally demanding. To address this, we propose a low-complexity approach using the \textit{c}ontrolled \textit{l}oosening-\textit{up} (CLuP) algorithm, which employs iterative refinement for efficient separation and estimation of radar targets. We achieve this through a nuclear norm restriction that stabilizes the problem. Numerical experiments demonstrate the robustness of this approach under high-mobility and noisy automotive environments, highlighting CLuP’s potential as a scalable, real-time solution for ISAC in future vehicular networks.
\end{abstract}

\section{Introduction}
\label{sec:intro}
Autonomous vehicles (AVs) are equipped with sophisticated sensing \cite{patole2017automotive} and communications \cite{gonzalez2025six} that enable them to navigate complex environments with minimal human intervention. Among these systems, radar plays a crucial role in detecting obstacles and determining the relative motion of objects, while communications channels facilitate data exchange with other vehicles and infrastructure \cite{bilik2019rise,sun2020mimo,ali2020leveraging}. Recently, integrated sensing and communications (ISAC) systems have emerged as a promising approach for spectrum sharing in AVs \cite{mishra2024signal}, wherein a shared use of resources by radar and communications also results in increased efficiency, reduced hardware complexity, and low latency \cite{MishraMulticarrier,MishraSpherical}.

Depending on the availability of information, ISAC systems may be blind or non-blind. In the former case, both radar and communications receivers lack knowledge of transmit signal and channel making this a very challenging problem. This scenario requires advanced techniques to overcome the complete absence of prior knowledge. Existing approaches leverage the sparsity of the channel and the low-dimensional subspace structure of waveforms for effective recovery. For instance, \cite{monsalve2024dual, monsalve2023beurling} proposed a nuclear norm-based optimization problem to enforce low-rank structure on the vectorized Hankel matrix that captures the unknown parameters and provides recovery guarantees grounded in Beurling-Selberg extremal function theory \cite{moitra2015super}. The \textit{dual-blind deconvolution} (DBD) introduced in \cite{vargas2023dual, vargas2022joint} exploits channel sparsity while minimizing the sum of multivariate atomic norms (SoMAN) \cite{mishra2015spectral}. This technique leverages positive hyperoctant polynomials \cite{dumitrescu2007positive} to derive a semidefinite program (SDP) formulation for SoMAN enabled estimation of unknown target and communications parameters. Subsequent works extend this to phased array receiver \cite{jacome2022multid} and graph theoretic approaches \cite{jacome2023factor}.

The non-blind \textit{dual-deconvolution} ISAC scenario is less restrictive but more commonly encountered in practice \cite{mishra2019toward}. Here, the transmitted radar waveform and estimates of communications channel are either known to the receiver or learned from data \cite{mishra2023next}. In this paper, we focus on this non-blind scenario, which enables more reliable data extraction and processing. Among prior works, \cite{masoud} employed alternating direction method of multipliers (ADMM) algorithm for joint recovery of communications message and radar target channel from overlaid signal vectors. However, their approach only focuses on the recovery of target reflectivity. While a recent study \cite{BothBlindNon} examines recovery guarantees in both blind and non-blind scenarios, leveraging the Fisher information matrix to establish performance bounds for target localization, it does not propose a method to efficiently extract the target parameters. In general, this is a computationally expensive task that generally relies on maximum likelihood (ML) estimation \cite{xu2023automotive}.

Contrary to these works, we propose a computationally-efficient technique for parameter estimation in automotive ISAC dual-deconvolution problems. In densely populated environments, radar signals often overlap with communications transmissions from other vehicles \cite{ali2020leveraging}. This overlap introduces challenges in distinguishing radar reflections from communications data and necessitates advanced signal processing to separate and accurately recover radar and communications information. We employ the recently proposed \textit{c}ontrolled \textit{l}oosening-\textit{up} (CLuP) based on random duality theory (RDT), as developed by Stojnic \cite{clup3, clup5}. CLuP is an optimization framework designed to handle complex problems by \textit{loosening} the constraints in a \textit{controlled} manner, thereby making them computationally tractable without significantly compromising solution quality. It solves a relaxed version of the original problem multiple times by minimizing the inner product between the solution from the previous iteration and the current one. The convergence guarantee is analyzed via RDT that leverages the statistical properties of random matrices and vectors to simplify complex constraints and objectives that arise in such problems. A key aspect of RDT is its reliance on duality principles, where a primal optimization problem is transformed into its dual counterpart, often resulting in a more tractable formulation. 

In the context of automotive ISAC, our CLuP-based algorithm exploits both the sparsity of target profiles or \textit{radar channel} and the low-dimensional subspace structure of communications messages, leading to enhanced estimation accuracy and computational efficiency. We modify CLuP for the dual-deconvolution problem by incorporating a nuclear norm restriction that exploits the sparsity of the radar channel, while assuming that the communications message resides within a low-dimensional subspace. To this end, we apply a vectorized Hankel lift operator to the radar channel parameters and subsequently impose a nuclear norm restriction on the resulting matrix. This formulation not only improves the convergence properties of the optimization but also ensures that the radar and communications components can be effectively decoupled and estimated with high accuracy. We evaluate the performance of the proposed method through extensive numerical experiments that demonstrate accurate recovery of communications messages and locations of targets. The inclusion of the nuclear norm restriction ensures robust recovery even in challenging operational settings such as the presence of noise. 

Throughout this paper, we use bold lowercase and bold uppercase letters to denote vectors (e.g., $\mathbf{x}$) and matrices (e.g., $\mathbf{X}$), respectively. The transpose and Hermitian (conjugate transpose) of a matrix $\mathbf{A}$ are $\mathbf{A}^T$ and $\mathbf{A}^H$, respectively. The notation $[\mathbf{a}]_i$ refers to the $i$-th element of vector $\mathbf{a}$; $[\mathbf{A}]_{j,k}$ denotes the element in the $j$-th row and $k$-th column of matrix $\mathbf{A}$; and $\mathbf{a}_i$ represents the $i$-th column of $\mathbf{A}$. The sets of real and complex numbers are $\mathbb{R}$ and $\mathbb{C}$, respectively. The notation diag($\mathbf{a}$) represents a diagonal matrix with elements of the vector $\mathbf{a}$ as its diagonal. The symbol $\delta(t)$ is the Dirac delta function and $\star$ denotes convolution operation. The Euclidean (nuclear) norm of vector $\mathbf{x}$ (matrix $\mathbf{X}$) is $\|\mathbf{x}\|$ ($\|\mathbf{X}\|_*$). We represent the inner product between vectors $\mathbf{x}$ and $\mathbf{y}$ by $\langle \mathbf{x},\mathbf{y}\rangle$. The canonical vector $\mathbf{e}_j$ is a unit vector with a $1$ in the $j$-th position and $0$ elsewhere. The notation $\textrm{Pr}(.)$ denotes the probability of its argument event.

\section{System model}
\label{sec:sysmod}
\begin{figure}[t]
    \centering
    \includegraphics[width=0.99\linewidth]{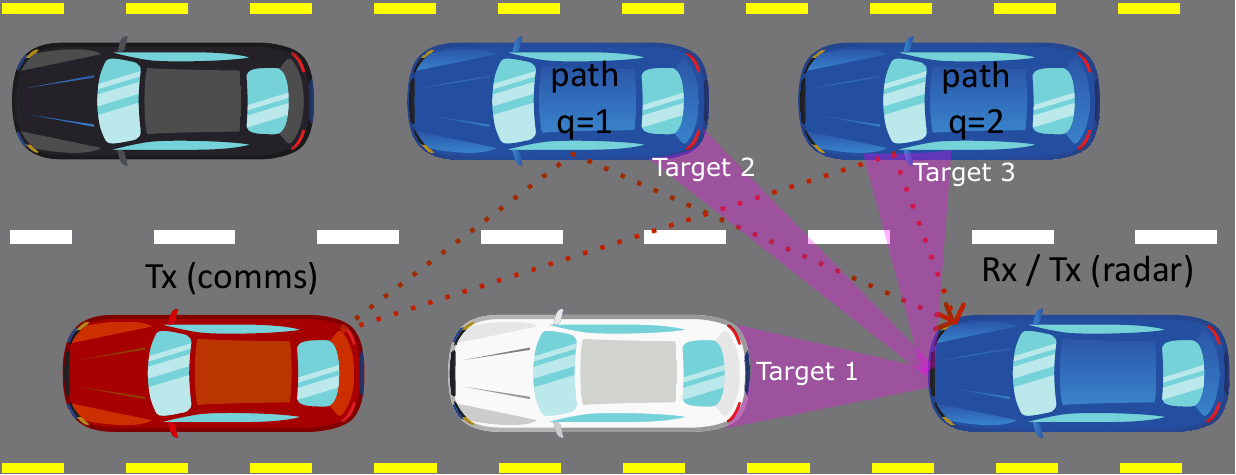}
    \caption{Illustration of the hybrid monostatic-bistatic setup of the joint radar-communications scenario. The radar transmitter (Tx) mounted on blue car on bottom right sends a signal to detect other vehicles (labeled as `Target 1', `Target 2' and so on) in its surroundings. It also acts a common receiver (Rx) for the radar signal reflected off from these targets and the communications message sent by the red car. }
    \label{fig:concept}
\end{figure}

Consider a hybrid monostatic-bistatic system \cite{dokhanchi2019mmWave,dokhanchi2019mono} in an automotive scenario (Figure \ref{fig:concept}). Here, a radar-equipped vehicle aims to detect nearby targets while simultaneously receiving communications signals from surrounding vehicles. The radar transmits a signal $s(t)$, assumed to be baseband here for simplicity but without loss of generality, periodically in a sequence of $P$ pulses with a pulse repetition interval (PRI) of $T$. The transmit pulse train is $x_r(t) = \sum_{p=1}^P s(t - pT)$. Assume $L$ targets on the scene, which are characterized by the vectors: targets' Swerling Model 1 \cite{skolnik2008radar} complex reflectivities  $\bsym{\beta} \in \mathbb{C}^L$, time delays $\bsym{\tau_r} \in \mathbb{R}^{L}$ (directly proportional to targets' ranges), and Doppler frequencies $\bsym{\nu_r} \in \mathbb{R}^L$ (directly proportion to target’s radial/Doppler velocity) \cite{rudresh2017finite}. The radar channel is
\begin{equation}
    h_r(t) = \sum_{\ell=1}^L [\bsym{\beta}]_{\ell} \delta(t - [\bsym{\tau_r}]_{\ell})e^{-2\mathrm{j}\pi [\bsym{\nu_r}]_{\ell} t},
\end{equation}
where $[\cdot]_l$ denotes the $l$-th element of the vector argument.

 The vehicular wireless communications system uses orthogonal frequency-division multiplexing (OFDM) to transmit $P$ baseband messages $x_p(t)$. These messages are distributed over $K$ sub-carriers, each with a symbol duration of $T$. The communicaions message is $x_p(t) = \sum_{k=0}^{K-1} [\mathbf{g}_p]_k e^{\mathrm{j}2\pi k \Delta f t}$, where $[\mathbf{g}_p]k$ represents the $p$-th complex symbol modulated onto the $k$-th sub-carrier frequency $f_k = k \Delta f$. The  transmit signal over $P$ frames is $x_c(t) = \sum_{p=0}^{P-1} x_p(t-pT)$. This signal propagates through $Q$ distinct paths characterized by the vectors of complex path gains $\bsym{\alpha} \in \mathbb{C}^Q$, time delays $\bsym{\tau_c} \in \mathbb{R}^Q$, and Doppler frequency shifts $\bsym{\nu_c} \in \mathbb{R}^Q$.The communications channel is
\begin{equation}
    h_c(t) = \sum_{q=1}^Q [\bsym{\alpha}]_{q} \delta(t - [\bsym{\tau_c}]_{q})e^{-2\mathrm{j}\pi [\bsym{\nu_c}]_{q} t}.
\end{equation}

The ISAC signal at the common receiver is the superposition of radar and communications signals convolved with their respective channels:
\begin{align}
    \tilde{y}(t) &= (h_r(t) \star x_r(t)) + (h_c(t) \star x_c(t)) + \xi(t) \nonumber\\
    &=\sum_{p=0}^{P-1}\sum_{\ell=0}^{L-1} [\bsym{\beta}]_{\ell} s(t - pT - [\bsym{\tau_r}]_{\ell}) e^{-2\mathrm{j}\pi [\bsym{\nu_r}]_{\ell} t} 
    \nonumber\\ 
    &+\sum_{p=0}^{P-1} \sum_{k=0}^{K-1}\sum_{q=0}^{Q-1} [\bsym{\alpha}]_{\ell} [\mathbf{g}_p]_k e^{\mathrm{j}2\pi k \Delta f (t-pT-[\bsym{\tau_c}]_{\ell})} e^{-2\mathrm{j}\pi [\bsym{\nu_c}]_{\ell} t} \nonumber\\ 
    &+ \xi(t),
\end{align}
where  $\xi(t)$ is additive white Gaussian noise\cite{vargas2023dual}. 
The duration $T$ is either known or estimated \cite{vargas2023dual}. To align the signals in time, define $y(t,p) = \tilde{y}(t + pT)$ such that
\begin{equation}
\begin{aligned}
    y(t, p) &=\sum_{\ell=0}^{L-1} [\bsym{\beta}]_{\ell} s(t - [\bsym{\tau_r}]_{\ell}) e^{-2\mathrm{j}\pi [\bsym{\nu_r}]_{\ell} t} 
    \\ 
    &+ \sum_{k=0}^{K-1}\sum_{q=0}^{Q-1} [\bsym{\alpha}]_{\ell} [\mathbf{g}_p]_k e^{\mathrm{j}2\pi k \Delta f (t-[\bsym{\tau_c}]_{\ell})} e^{-2\mathrm{j}\pi [\bsym{\nu_c}]_{\ell} t} + \xi(t,p).
\end{aligned}
\label{eq:y1}
\end{equation}
Note that $y(t, 0)$ and the shifted signal $y(t, p)$ share the same set of parameters. The phase rotation for each radar pulse and message remains constant during the period $T$, so $e^{-2\mathrm{j}\pi [\boldsymbol{\nu_r}]_{\ell} t} \approx e^{-2\mathrm{j}\pi [\boldsymbol{\nu_r}]_{\ell} pT}$ and $e^{-2\mathrm{j}\pi [\boldsymbol{\nu_c}]_{\ell} t} \approx e^{-2\mathrm{j}\pi [\boldsymbol{\nu_c}]_{\ell} pT}$ \cite{monsalve2023beurling}. We then compute the 1-D continuous-time Fourier transform (CTFT) of \eqref{eq:y1} along the time dimension, with a frequency range of $f \in [-B/2, B/2]$, where $B$ denotes the effective bandwidth for both radar and communications signals. The frequency-domain samples have OFDM spacing, namely, $f_n = \frac{B n}{M} = n \Delta f$, where $n = -N, \ldots, N$ and $M = 2N + 1$. The CTFT is
\begin{equation}
    \begin{aligned}
        Y(f_n, p) &= \sum_{\ell=0}^{L-1} [\bsym{\beta}]_{\ell} S(f_n) e^{-2\mathrm{j}\pi (n\Delta f[\bsym{\tau_r}]_{\ell}+[\bsym{\nu_r}]_{\ell} pT)}
        \\&+ \sum_{q=0}^{Q-1} [\bsym{\alpha}]_{q} [\mathbf{g}_p]_n e^{-2\mathrm{j}\pi (n\Delta f[\bsym{\tau_c}]_{q}+[\bsym{\nu_c}]_{q} pT)} + \xi(f_n, p),
    \end{aligned}
    \label{eq:problem1}
\end{equation}
where $S(f_n)$ is the CTFT of $s(t)$.

To make the parameters independent of the channel-specific settings, we normalize the channel parameters as $\bsym{\tilde{\tau}_r} = \bsym{\tau_r} / T \in [0,1]^{L}$, $\bsym{\tilde{\tau}_c} = \bsym{\tau_c} / T \in [0,1]^{Q}$, $\bsym{\tilde{\nu}_r} = \bsym{\nu_r} / \Delta f \in [0,1]^{L}$, $\bsym{\tilde{\nu}_c} = \bsym{\nu_c} / \Delta f \in [0,1]^{Q}$. Define $[\mathbf{s}]_n$ as the samples of the CTFT $S(f)$ taken at frequencies $f_n$, and $[\mathbf{g}]_m = [\mathbf{g}_p]_n$ with $m = n+N + pM$. Stacking the measurements in a vector $[\mathbf{y}] = Y(n,p)$ and rewriting the overlaid receiver signal in \eqref{eq:problem1} as a linear system gives
    \begin{align}
        \mathbf{y} 
            =\text{diag}(\mathbf{s})\mathbf{V_r}\bm{\beta} +  \text{diag}(\mathbf{g})\mathbf{V_c}\bm{\alpha} +\bsym{\xi}\label{eq:problemjoint},
    \end{align}
where $\mathbf{V_r}$ and $\mathbf{V_c}$ are Vandermonde matrices that collect the delay and Doppler parameters of the channels. In this traditional non-blind set-up, the radar waveform $\mathbf{s}$ and communications channel $\mathbf{V}_c$ are known. Our objective is to recover the radar channel parameters -- $\bsym{\tilde{\tau}_r}, \bsym{\tilde{\nu}_r}$ and $\beta$ -- and the communications transmit message $\mathbf{g}$.

\section{CLuP-Based Dual-Deconvolution}
In order to retrieve the unknown parameters from the dual-deconvolution model in \eqref{eq:problemjoint}, we collect all the unknowns for both radar and communications in, respectively, the following vectors
\begin{align}
    \mathbf{h_r}_p = \sum_{\ell=0}^{L-1} [\bsym{\beta}]_{\ell} \mathbf{a}([\bsym{\tau_r}]_{\ell})e^{-2\mathrm{j}\pi [\bsym{\nu_r}]_p}, \label{eq_hr}
\end{align}
and
\begin{align}    
    \mathbf{h_c}_p = \sum_{q=0}^{Q-1} [\bsym{\alpha}]_{q} \mathbf{a}([\bsym{\tau_c}]_{q})e^{-2\mathrm{j}\pi [\bsym{\nu_c}]_p},
    \label{eq_hc}
\end{align}
with $\alpha= e^{-\mathrm{j}2\pi}$, and  $\mathbf{a}([\bsym{\tau}]_{\ell})=[\alpha^{[\bsym{\tau}]_{\ell}\times 0},\alpha^{[\bsym{\tau}]_{\ell}\times 1}, \ldots, \alpha^{[\bsym{\tau}]_{\ell}\times M-1}]^T \in \mathbb{C}^{M}$. Note that \eqref{eq_hr}, being a collection of radar target's delay, Doppler, and gain parameters, is unknown to the common receiver. We represent the communications messages in a low-dimensional subspace as \cite{lee2016blind,li2019rapid,ahmed2013blind,kuo2019geometry,ahmed2018leveraging}.
\begin{equation}
    \mathbf{g}= \mathbf{Dv}=\text{diag}(\mathbf{D}_0,\ldots,\mathbf{D}_P)[\mathbf{v}_0,\ldots, \mathbf{v}_P]^T,    
\end{equation}
where each $\mathbf{D}_p \in \mathbb{R}^{M\times J}$, $\mathbf{v}_p \in \mathbb{R}^{J}$ with $J$ is the subspace dimension such that $J<M$. In our dual-deconvolution scenario, the vectors $\mathbf{h_r}_p$ and $\mathbf{v}_p$ are unknown. 

Define $\mathbf{h_r} = \begin{bmatrix}
        \mathbf{h_r}_0^T & \mathbf{h_r}_1^T & \ldots & \mathbf{h_r}_P^T\\
    \end{bmatrix}^T  \in \mathbb{R}^{MP}$ and $\mathbf{v} = \begin{bmatrix}
        \mathbf{v}_0^T & \mathbf{v}_1^T & \ldots & \mathbf{v}_P^T\\
    \end{bmatrix}^T \in \mathbb{R}^{JP}$. Collect all unknown quantities in the the following vector 
\begin{align}
     \mathbf{x} = \begin{bmatrix}
        \mathbf{h_r}^T & \mathbf{v}^T \\
    \end{bmatrix}^T \in \mathbb{R}^{(J+M)P}.
    \label{eq:xdef}
\end{align}

Define $\mathbf{y}_r=\mathcal{A}_r(\mathbf{h_r})$, where $\mathcal{A}_r$ is a linear operator defined as
\begin{equation}
    [\mathcal{A}_r(\mathbf{h})]_i = \langle \mathbf{s}_i \mathbf{e}_i , \mathbf{h} \rangle.
\end{equation}
Similarly, for the communications part, \textit{mutatis mutandis} we construct the linear operator $\mathcal{A}_c$ that contains the channel parameters and is convolved with the message as $\mathbf{y}_c = \mathcal{A}_c(\mathbf{v})$ such that
\begin{equation}
    [\mathcal{A}_c(\mathbf{v})]_j = \langle \mathbf{d}_j \mathbf{e}_j^T \mathbf{h}_c, \mathbf{v}
    \rangle.
\end{equation}

By defining $\mathcal{A}(\mathbf{x}) = \mathcal{A}_r(\mathbf{h_r}) + \mathcal{A}_c(\mathbf{v})$, the overlaid received signal becomes 
\begin{equation}
    \mathbf{y} = \mathcal{A}_r(\mathbf{h}) + \mathcal{A}_c(\mathbf{v}) + \bsym{\xi} = \mathcal{A}(\mathbf{x}) + \bsym{\xi}.
\end{equation}

For the sake of simplicity but without loss of generality, the operating
conditions about the target parameters and communications messages include
\begin{description}
    \item[A1] ``Relative reflectivities": The target reflectivies are normalized by their maximum values such that they sum to one, i.e., $\sum_{\ell=^1}^{L} [\bsym{\beta}]_{\ell} \leq 1$.

    \item[A2] ``Unit-norm message": The communications message vector has a unit norm, i.e., \( \| \mathbf{v} \| = 1 \). 

\end{description}
    
From the definition of $\mathbf{h_r}_p$ in \eqref{eq_hr} and conditions \textbf{A1}-\textbf{A2}, it follows that $\text{max}_{i} |[\mathbf{x}]_i|\leq 1$. Denote the normalized vector of unknowns as $\mathbf{\tilde{x}} = (1/\sqrt{N}) \mathbf{x}$, with $N=(M+J)\times P$. This implies that $\|\mathbf{\tilde{x}}\|\leq 1$ such that the dual-deconvolution measurements become
\begin{equation}
     \mathbf{y}  = \sqrt{N} \mathcal{A}(\mathbf{\tilde{x}}) + \bsym{\xi}. \label{eq:mimo_ml}
\end{equation}
This is a classical multiple-input multiple-output (MIMO) ML problem that we solve using CLuP. 

Recall the CLuP optimization problem \cite{clup3, clup5}, wherein initializing with a random vector $\mathbf{x}^{(0)}$, the goal is to iteratively solve 
\begin{align}
\mathbf{\hat{x}}^{(i+1)} = \arg \min_{\mathbf{x}} & \, -(\mathbf{x}^{(i)})^H \mathbf{x}\nonumber \\
\text{subject to} & \, \| \mathbf{y} - \mathcal{A}( \mathbf{x}) \| \leq r, \nonumber\\
& \mathbf{x} \in \mathcal{X}, \quad
 \psi(\mathbf{x}), \label{eq:clup1} 
\end{align}
where $\mathcal{X} = \{-1/\sqrt{N},1/\sqrt{N}\}^{N}$ is the ``box'' constraint, $\psi(\mathbf{x})$ is any convex restriction, and $r$ is the radius parameter. In CLuP, choice of $r$ is critical for exact recovery. The CLuP procedure involved solving \eqref{eq:clup1} multiple times using an off-the-shelf solver such as a splitting conic solver \cite{scs}. In each iteration, the solution $\mathbf{\hat{x}}^{(i+1)}$ obtained in the last iteration is used for the next iteration by replacing $\mathbf{x}^{(i)}$ by the normalized vector 
\begin{equation}
    \mathbf{x}^{(i+1)} = \frac{\mathbf{\hat{x}}^{(i+1)}}{\|\mathbf{\hat{x}}^{(i+1)}\|}.
\end{equation}

To simplify the solution of \eqref{eq:clup1}, recall that the number of targets is much smaller than the number of measurements, i.e., $L \ll M$. Then, we construct a low-rank Hankel matrix by setting with $P + 1= p_1 + p_2$ using the operator
\begin{equation}
    \mathcal{H}(\mathbf{h_r}) = \begin{bmatrix}
       \mathbf{h_r}_{0} & \mathbf{h_r}_{1} & \ldots &  \mathbf{h_r}_{p_2-1} \\
       \mathbf{h_r}_{1} & \mathbf{h_r}_{2} & \ldots &  \mathbf{h_r}_{p_2}\\
       \vdots  & \vdots & \ddots & \vdots \\
        \mathbf{h_r}_{p_1-1} & \mathbf{h_r}_{p_1} & \ldots &  \mathbf{h_r}_{P-1} \\
    \end{bmatrix} \in \mathbb{R}^{(p_1M) \times p_2}.\label{eq:hankel}
\end{equation}
When some separation conditions are satisfied on $\tau_r$ and $\nu_r$, the $\text{rank}(\mathcal{H}(\mathbf{h}))=L$\cite{monsalve2024dual}. Consequently, if $L<p_2$, the resulting Hankel matrix is low-rank\cite{monsalve2023beurling, monsalve2024dual}. This leads to a convex restriction to \eqref{eq:clup1} as 
\begin{align}
\mathbf{\hat{x}}^{(i+1)} = \arg \min_{\mathbf{x}} & \, -(\mathbf{x}^{(i)})^T \mathbf{x}\nonumber \\
\text{subject to} & \, \| \mathbf{y} - \mathcal{A}( \mathbf{x}) \| \leq r,\label{eq:clup2}   \nonumber\\
& \|\mathcal{H}(\mathbf{x})\|_*\leq s, \nonumber\\
& \mathbf{x} \in \left[\frac{-1}{\sqrt{N}}, \frac{-1}{\sqrt{N}}\right]^N.
\end{align}
The box constraint is applied to both real and imaginary parts of $\mathbf{x}$. That is, if $\mathbf{x}=\mathbf{x}_R+\mathrm{j}\mathbf{x}_I$, then the constraint applies component-wise to both $\mathbf{x}_R$ and $\mathbf{x}_I$. 

Solution to problem \eqref{eq:clup2} converges to the exact point $\mathbf{\tilde{x}}$ if the radius $r$ is selected accurately. To this end, we set $r = c_0 r_{m}$ such that $c_0$ is a constant and $r_{m}$ is the minimum possible radius found by solving the polytope relaxation 
\begin{align}
r_{m} = \arg \min_{\mathbf{x}} & \, \frac{1}{\sqrt{N}} \| \mathbf{y} - \mathcal{A} (\mathbf{x}) \| \nonumber\\
\text{subject to} & \, \mathbf{x} \in \left[\frac{-1}{\sqrt{N}}, \frac{-1}{\sqrt{N}}\right]^N.
\end{align}
The parameter $s=c_1 s_{m}$, with $c_1$ as a constant, is found by solving 
\begin{align}
s_{m} = \arg \min_{\mathbf{x}} & \, \|\mathcal{H}(\mathbf{x})\|_* \nonumber\\
\text{subject to} & \, \| \mathbf{y} - \mathcal{A}( \mathbf{x}) \| \leq r,\label{eq:clup3} \nonumber \\
& \mathbf{x} \in \left[\frac{-1}{\sqrt{N}}, \frac{-1}{\sqrt{N}}\right]^N.
\end{align}

We provide the following Proposition~\ref{prop:conv} to show that solving \eqref{eq:clup2} via CLuP converges to the exact point with high probability.
\begin{proposition}
\label{prop:conv}
    Assume the elements of $\mathbf{D}$, $\mathbf{h}_c$ and $\mathbf{s}$ are independent and identically distributed elements drawn from a standard normal distribution and the conditions \textbf{A1}-\textbf{A2} hold, then \eqref{eq:clup2} converges to the exact point with high probability as $N\to \infty$. 
\end{proposition}
\begin{IEEEproof}
As the sequence of solutions of \eqref{eq:clup2} converges to the optimal solution, we can rewrite it as
\begin{align}
\minimize_{\mathbf{x}} & \, \|\mathbf{x}\|\nonumber \\
\text{subject to} & \, \| \mathbf{y} - \mathcal{A}( \mathbf{x}) \| \leq r,\nonumber   \\
\quad & \|\mathcal{H}(\mathbf{x})\|_*\leq s, \nonumber\\
& \mathbf{x} \in \left[\frac{-1}{\sqrt{N}}, \frac{-1}{\sqrt{N}}\right]^N.\label{eq:clup4}
\end{align}
The conditions \textbf{A1}-\textbf{A2} imply that the absolute value of each element of $\mathbf{\tilde{x}}$ is $1/\sqrt{N}$. Denote $n_1 = \|\mathbf{x}\|$ and $n_2 = \mathbf{\tilde{x}}^H \mathbf{x}$. Then, by setting $\bsym{\xi} = \sigma \mathbf{n}$ with $\mathbf{n}$ a random vector from a standard distribution, \eqref{eq:clup4} becomes 
\begin{align}
\minimize_{\mathbf{x}} & \, \|\mathbf{x}\|\nonumber \\
\text{subject to} & \, \left\| \mathcal{A}(\mathbf{\tilde{x}} - \mathbf{x}) + \sigma \mathbf{n} \right\| \leq r,\nonumber   \\
\quad & \|\mathcal{H}(\mathbf{x})\|_*\leq s, \nonumber\\
& \mathbf{x} \in \left[\frac{-1}{\sqrt{N}}, \frac{-1}{\sqrt{N}}\right]^N,\label{eq:clup5}
\end{align}
where $\sigma^2$ is the noise variance. The Lagrange dual of \eqref{eq:clup5} is \cite{clup5} 
\begin{align}
\bsym{\eta}(\mathcal{A}, \mathcal{H}) = \max_{\stackrel{\gamma_i\geq 0}{ i=1,2}} \min_{\mathbf{x}} \max_{\|\bsym{\lambda}\|=1} & \, \|\mathbf{x}\| +\gamma_1 \bsym{\lambda}^T (\mathcal{A}(\mathbf{\tilde{x}} - \mathbf{x}) + \sigma \mathbf{n})\nonumber\\
   &- \gamma_1 r +  \gamma_2\left(\sum_{i=0}^{p_2}\rho_i\left(\mathcal{H}(\mathbf{x})\right) - s\right), \nonumber\\
\text{subject to}  & \; \mathbf{x} \in \left[\frac{-1}{\sqrt{N}}, \frac{-1}{\sqrt{N}}\right]^N,\label{eq:clup6}
\end{align}
where $\rho_i(\cdot)$ gives the $i$-th singular value of its argument; $\bsym{\lambda}$ and $\gamma_i$, $i=1,2$, are the Lagrange multipliers. From \cite[Eq. (45)]{clup6} that provides a general structure of a Lagrange dual subjected to a linear set of constraints as the one described in \eqref{eq:clup6} and invoking \cite[Lemmata 2 and 4]{clup6} that provide guarantees for lower and upper bounds on the value of the objective function in \eqref{eq:clup6}, it follows that there exist vector constants $\bsym{\mu}^{(l)}$ and $\bsym{\mu}^{(u)}$ such that 
\begin{equation}
    \lim_{N \to \infty} \textrm{Pr}(\bsym{\mu}^{(u)} > \bsym{\eta}(\mathcal{A}, \mathcal{H}) >\bsym{\mu}^{(l)} ) = 1.
\end{equation}
This completes the proof. 
\end{IEEEproof}
Once CLuP recovers the vector $\mathbf{\tilde{x}}$ by solving \eqref{eq:clup2}, we construct the Hankel matrix $\mathcal{H}(\mathbf{\tilde{x}})$ given in \eqref{eq:hankel}. Then, we apply a spectral recovery algorithm, such as MUSIC \cite{zhang2009music} or ESPRIT \cite{roy1989esprit} on the resulting Hankel matrix to estimate $\tau_r$ and $\nu_r$. The messages are estimated as $\mathbf{g}= \mathbf{D\tilde{v}}$ where $\mathbf{\tilde{v}}$ is extracted from $\mathbf{\tilde{x}}$ following \eqref{eq:xdef}. 

\section{Numerical Experiments}
\begin{figure}[ht]
    \centering
    \includegraphics[width=0.9\linewidth]{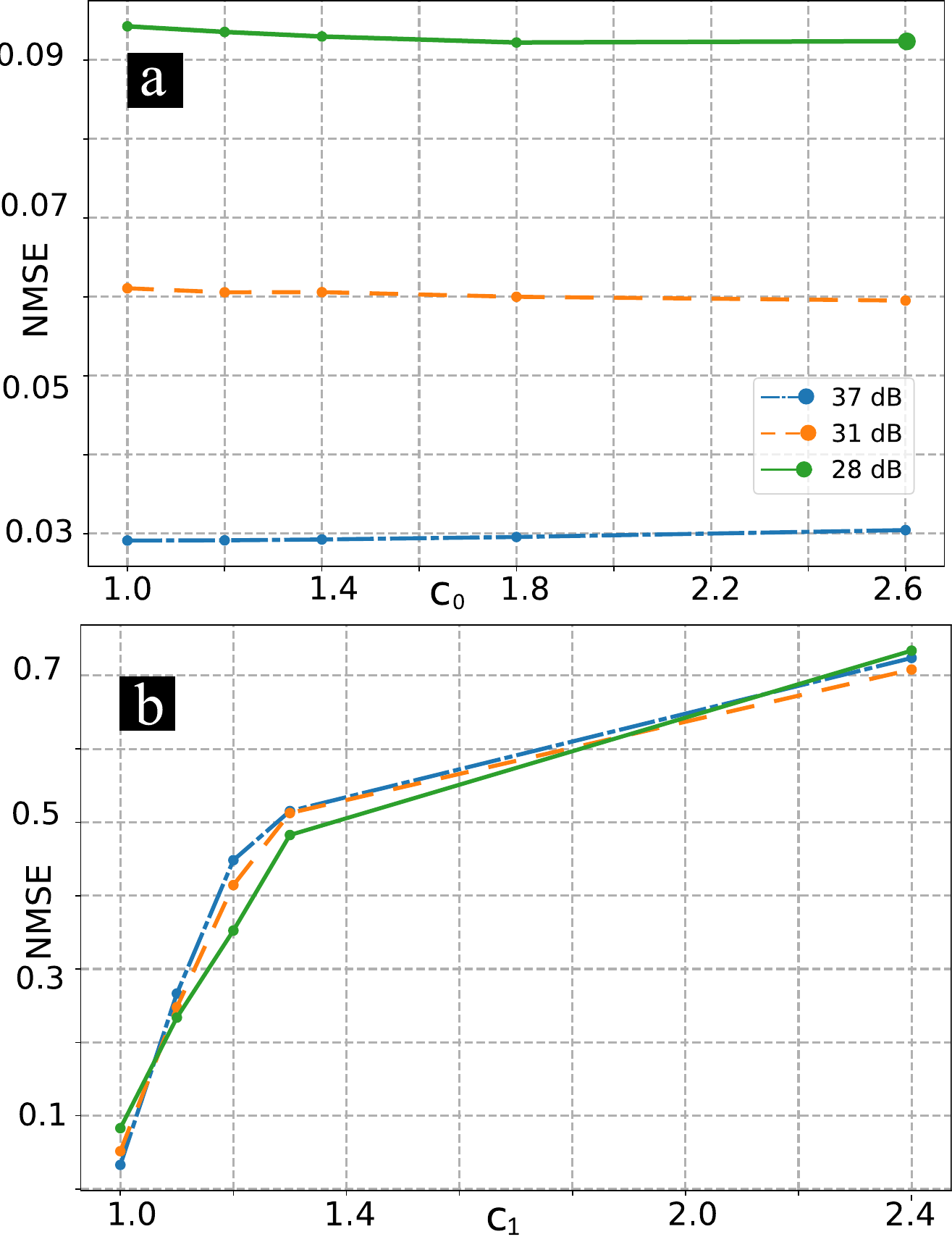}
    \caption{NMSE in recovering $\mathbf{x}$ for SNR $=37$, $31$, and $28$ dB with varying values of CLuP parameters (a) $c_0$ and (b) $c_1$.}
    \label{fig:movingr}
\end{figure}
\begin{figure*}
    \centering
    \includegraphics[width=1.0\linewidth]{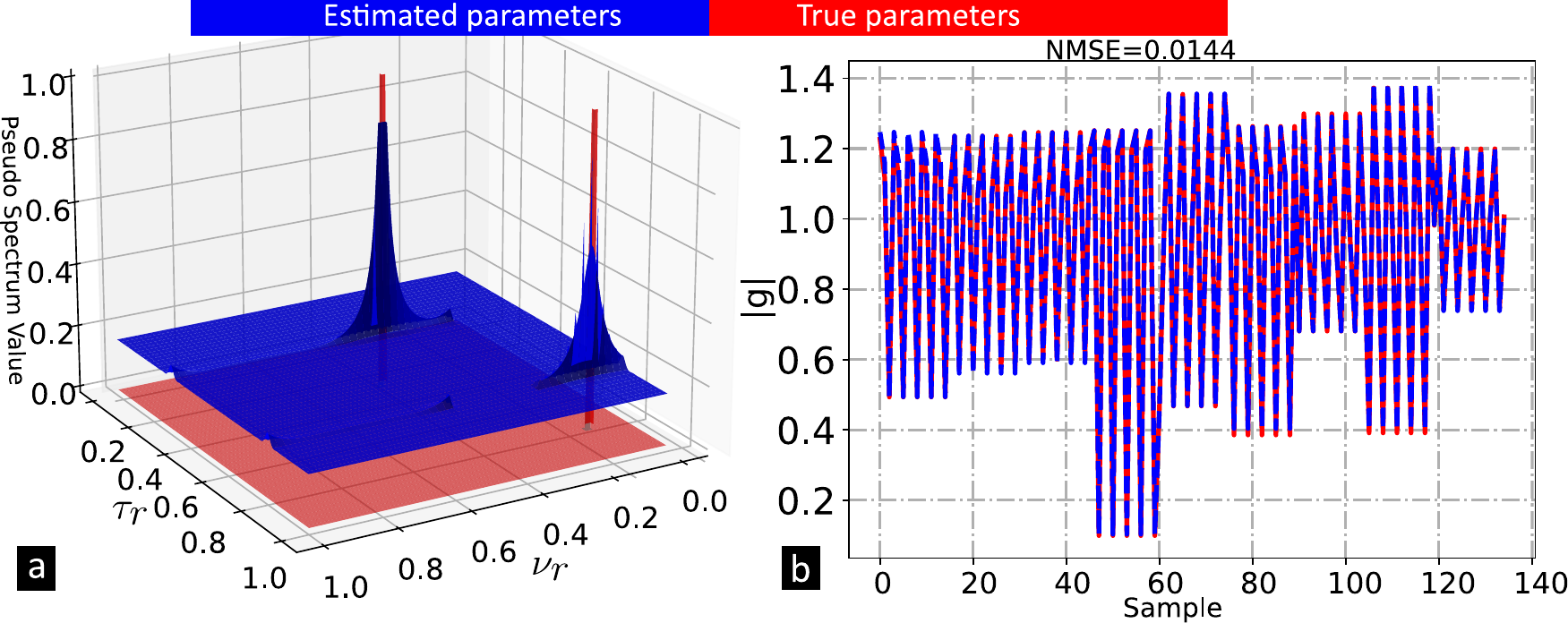}
    \caption{(a) CLuP recovered pseudo-spectrum of radar channel compared with true target parameter values for $L=2$ (b) Estimated norm of communications message vector compared with true values for J=2.}
    \label{fig:specific}
\end{figure*}
\begin{figure}[ht]
    \centering
    \includegraphics[width=0.99\linewidth]{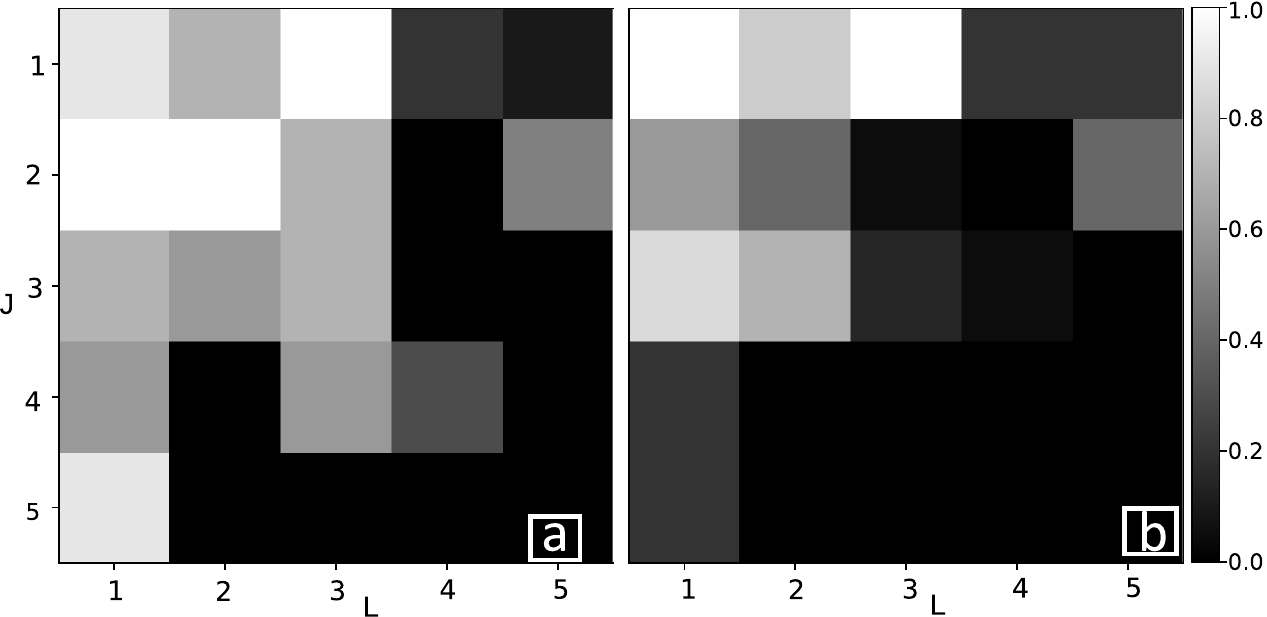}
    \caption{The probability of successful reconstruction, averaged over 20 trials, with $P = 9$ and $M = 11$ for dual-deconvolution recovery using (a) CLuP and (b) ADMM.}
    \label{fig:probrecons}
\end{figure}
We validated our approach through numerical experiments. First, we tested CLuP's performance by varying the parameters $s$ and $r$ gradually approaching their actual values. We set $L=2$, $J=2$, $M=15$, and $P=9$. The channel parameters $\tau_r$, $\tau_c,$ $\nu_r$, and $\nu_c$ were drawn from a uniform distribution over $[0,1]$. We used normalized mean squared error, NMSE $= \|\mathbf{x}-\mathbf{\tilde{x}}\| /\|\mathbf{x}\|$  as the recovery performance benchmark. Define the signal-to-noise ratio, SNR $= 10 \log_{10} \left( \frac{\|\mathbf{x}\|^2}{\|\sigma\mathbf{n}\|^2} \right)$. 

Figure~\ref{fig:movingr}a shows the impact of the parameter $r=c_0r_{m}$. In low noise regimes, setting $c_0=1$ yields better (lower) NMSE values. Higher noise levels require higher values of $c_0$. For instance, at SNR $=28$ dB, setting $c_0=1.8$ produces better estimation results. Figure~\ref{fig:movingr}b shows the NMSE performance with the increasing values of the parameter $c_1$. It follows that the algorithm is more consistent with the SNR variation for this parameter.

For $L=J=2$, the pseudo-spectrum of the estimated radar channel computed using MUSIC in Figure \ref{fig:specific}a shows the exact recovery of target parameters $\bsym{\tau_r}$ and $\bsym{\nu_r}$. Figure~\ref{fig:specific}b similarly shows an accurate estimation of the communications message vector. 

Finally, we compare CLuP against a modified version of the ADMM-based dual-deconvolution of \cite{masoud}. Figures~\ref{fig:probrecons}a and b show the probability of successful reconstruction of the vector $\mathbf{x}$ averaged over $20$ trials for different values of the communications channel subspace dimension $J$ and the number of radar targets $L=1$ for CLuP and ADMM, respectively. Here, successful reconstruction implies NMSE $<0.1$. Note that the approach in \cite{masoud} only recovers reflectivities. Hence, we also adapted that method to recover other target parameters.

\section{Summary}
Dual-deconvolution has a rich heritage of research in signal processing. While techniques such as morphological component analysis \cite{starck2010sparse}, successive interference cancellation \cite{patel2002analysis} and ADMM \cite{masoud} have previously been employed to address various aspects of this problem. However, these methods ignored the exact recovery of parameters in a more general setting such as automotive ISAC. In this paper, we introduced a novel CLuP-based recovery algorithm to solve this challenging optimization problem. The method balances computational efficiency and solution accuracy by jointly harnessing the common structures of radar channels and communications messages. A key innovation of our approach to CLuP is imposing a nuclear norm constraint that enforces low-rank approximations in the vectorized Hankel-lifted representation of radar channel parameters. This leads to enhanced ability of our algorithm to converge robustly to reliable solutions.
\bibliographystyle{IEEEtran}
\bibliography{main}

\begin{thebibliography}{10}
\providecommand{\url}[1]{#1}
\csname url@samestyle\endcsname
\providecommand{\newblock}{\relax}
\providecommand{\bibinfo}[2]{#2}
\providecommand{\BIBentrySTDinterwordspacing}{\spaceskip=0pt\relax}
\providecommand{\BIBentryALTinterwordstretchfactor}{4}
\providecommand{\BIBentryALTinterwordspacing}{\spaceskip=\fontdimen2\font plus
\BIBentryALTinterwordstretchfactor\fontdimen3\font minus \fontdimen4\font\relax}
\providecommand{\BIBforeignlanguage}[2]{{%
\expandafter\ifx\csname l@#1\endcsname\relax
\typeout{** WARNING: IEEEtran.bst: No hyphenation pattern has been}%
\typeout{** loaded for the language `#1'. Using the pattern for}%
\typeout{** the default language instead.}%
\else
\language=\csname l@#1\endcsname
\fi
#2}}
\providecommand{\BIBdecl}{\relax}
\BIBdecl

\bibitem{patole2017automotive}
S.~M. Patole, M.~Torlak, D.~Wang, and M.~Ali, ``Automotive radars: {A} review of signal processing techniques,'' \emph{IEEE Signal Processing Magazine}, vol.~34, no.~2, pp. 22--35, 2017.

\bibitem{gonzalez2025six}
N.~Gonz{\'a}lez-Prelcic, D.~Tagliaferri, M.~F. Keskin, H.~Wymeersch, and L.~Song, ``Six integration avenues for {ISAC} in {6G} and beyond: {A} forward-looking vision,'' \emph{IEEE Vehicular Technology Magazine}, vol.~20, no.~1, pp. 18--39, 2025.

\bibitem{bilik2019rise}
I.~Bilik, O.~Longman, S.~Villeval, and J.~Tabrikian, ``The rise of radar for autonomous vehicles: {S}ignal processing solutions and future research directions,'' \emph{IEEE signal processing Magazine}, vol.~36, no.~5, pp. 20--31, 2019.

\bibitem{sun2020mimo}
S.~Sun, A.~P. Petropulu, and H.~V. Poor, ``{MIMO} radar for advanced driver-assistance systems and autonomous driving: {A}dvantages and challenges,'' \emph{IEEE Signal Processing Magazine}, vol.~37, no.~4, pp. 98--117, 2020.

\bibitem{ali2020leveraging}
A.~Ali, N.~Gonzalez-Prelcic, R.~W. Heath, and A.~Ghosh, ``Leveraging sensing at the infrastructure for {mmWave} communication,'' \emph{IEEE Communications Magazine}, vol.~58, no.~7, pp. 84--89, 2020.

\bibitem{mishra2024signal}
K.~V. Mishra, M.~R.~B. Shankar, B.~Ottersten, and A.~L. Swindlehurst, \emph{Signal processing for joint radar-communications}.\hskip 1em plus 0.5em minus 0.4em\relax Wiley-IEEE Press, 2024.

\bibitem{MishraMulticarrier}
S.~Bhattacharjee, K.~V. Mishra, R.~Annavajjala, and C.~R. Murthy, ``Multi-carrier wideband {OCDM}-based {THz} automotive radar,'' in \emph{IEEE International Conference on Acoustics, Speech and Signal Processing}, 2023, pp. 1--5.

\bibitem{MishraSpherical}
A.~M. Elbir, K.~V. Mishra, and S.~Chatzinotas, ``Spherical wavefront near-field {DoA} estimation in {THz} automotive radar,'' in \emph{European Conference on Antennas and Propagation}, 2024, pp. 1--5.

\bibitem{monsalve2024dual}
J.~Monsalve, E.~Vargas, K.~V. Mishra, B.~M. Sadler, and H.~Arguello, ``Dual-blind deconvolution in {ISAC} receiver using multi-dimensional {B}eurling-{S}elberg functions,'' in \emph{IEEE Radar Conference}, 2024, pp. 1--6.

\bibitem{monsalve2023beurling}
------, ``{B}eurling-{S}elberg extremization for dual-blind deconvolution recovery in joint radar-communications,'' in \emph{IEEE International Workshop on Computational Advances in Multi-Sensor Adaptive Processing}, 2023, pp. 246--250.

\bibitem{moitra2015super}
A.~Moitra, ``Super-resolution, extremal functions and the condition number of {V}andermonde matrices,'' in \emph{Annual ACM Symposium on Theory of Computing}, 2015, pp. 821--830.

\bibitem{vargas2023dual}
E.~Vargas, K.~V. Mishra, R.~Jacome, B.~M. Sadler, and H.~Arguello, ``Dual-blind deconvolution for overlaid radar-communications systems,'' \emph{IEEE Journal on Selected Areas in Information Theory}, vol.~4, pp. 75--93, 2023.

\bibitem{vargas2022joint}
------, ``Joint radar-communications processing from a dual-blind deconvolution perspective,'' in \emph{IEEE International Conference on Acoustics, Speech and Signal Processing}, 2022, pp. 5622--5626.

\bibitem{mishra2015spectral}
K.~V. Mishra, M.~Cho, A.~Kruger, and W.~Xu, ``Spectral super-resolution with prior knowledge,'' \emph{IEEE Transactions on Signal Processing}, vol.~63, no.~20, pp. 5342--5357, 2015.

\bibitem{dumitrescu2007positive}
B.~Dumitrescu, \emph{Positive trigonometric polynomials and signal processing applications}.\hskip 1em plus 0.5em minus 0.4em\relax Springer, 2007.

\bibitem{jacome2022multid}
R.~Jacome, K.~V. Mishra, E.~Vargas, B.~M. Sadler, and H.~Arguello, ``Multi-dimensional dual-blind deconvolution approach toward joint radar-communications,'' in \emph{IEEE International Workshop on Signal Processing Advances in Wireless Communication}, 2022, pp. 1--5.

\bibitem{jacome2023factor}
------, ``Factor graph expectation maximization dual-blind deconvolution for dynamic integrated sensing and communications,'' in \emph{IEEE International Workshop on Computational Advances in Multi-Sensor Adaptive Processing}, 2023, pp. 1--5.

\bibitem{mishra2019toward}
K.~V. Mishra, M.~R.~B. Shankar, V.~Koivunen, B.~Ottersten, and S.~A. Vorobyov, ``Toward millimeter-wave joint radar communications: {A} signal processing perspective,'' \emph{IEEE Signal Processing Magazine}, vol.~36, no.~5, pp. 100--114, 2019.

\bibitem{mishra2023next}
K.~V. Mishra, M.~R.~B. Shankar, and M.~Rangaswamy, \emph{Next-generation cognitive radar systems}.\hskip 1em plus 0.5em minus 0.4em\relax IET Press, 2023.

\bibitem{masoud}
M.~Farshchian and I.~Selesnick, ``A dual-deconvolution algorithm for radar and communication black-space spectrum sharing,'' in \emph{International Workshop on Compressed Sensing Theory and its Applications to Radar, Sonar and Remote Sensing}, 2016, pp. 6--10.

\bibitem{BothBlindNon}
N.~Zhao, Q.~Chang, X.~Shen, Y.~Wang, and Y.~Shen, ``Joint target localization and data detection in bistatic {ISAC} networks,'' \emph{IEEE Transactions on Communications}, pp. 1--1, 2024.

\bibitem{xu2023automotive}
L.~Xu, S.~Sun, K.~V. Mishra, and Y.~D. Zhang, ``Automotive {FMCW} radar with difference co-chirps,'' \emph{IEEE Transactions on Aerospace and Electronic Systems}, vol.~59, no.~6, pp. 8145--8165, 2023.

\bibitem{clup3}
M.~Stojnic, ``Controlled loosening-up ({CLuP}) – achieving exact {MIMO} {ML} in polynomial time,'' \emph{arXiv preprint arXiv:1909.01175}, 2019.

\bibitem{clup5}
------, ``Rephased {CLuP},'' \emph{arXiv preprint arXiv:2011.11527}, 2020.

\bibitem{dokhanchi2019mmWave}
S.~H. Dokhanchi, M.~R.~B. Shankar, K.~V. Mishra, and B.~Ottersten, ``A {mmWave} automotive joint radar-communications system,'' \emph{IEEE Transactions on Aerospace and Electronic Systems}, vol.~55, no.~3, pp. 1241--1260, 2019.

\bibitem{dokhanchi2019mono}
S.~H. Dokhanchi, M.~Alaee-Kerahroodi, M.~R.~B. Shankar, and B.~Ottersten, ``Mono-static automotive joint radar-communications system,'' in \emph{IEEE Annual International Symposium on Personal, Indoor and Mobile Radio Communications}, 2019, pp. 1--6.

\bibitem{skolnik2008radar}
M.~I. Skolnik, \emph{Radar handbook}, 3rd~ed.\hskip 1em plus 0.5em minus 0.4em\relax McGraw-Hill, 2008.

\bibitem{rudresh2017finite}
S.~Rudresh and C.~S. Seelamantula, ``Finite-rate-of-innovation-sampling-based super-resolution radar imaging,'' \emph{IEEE Transactions on Signal Processing}, vol.~65, no.~19, pp. 5021--5033, 2017.

\bibitem{lee2016blind}
K.~Lee, Y.~Li, M.~Junge, and Y.~Bresler, ``Blind recovery of sparse signals from subsampled convolution,'' \emph{IEEE Transactions on Information Theory}, vol.~63, no.~2, pp. 802--821, 2016.

\bibitem{li2019rapid}
X.~Li, S.~Ling, T.~Strohmer, and K.~Wei, ``Rapid, robust, and reliable blind deconvolution via nonconvex optimization,'' \emph{Applied and Computational Harmonic Analysis}, vol.~47, no.~3, pp. 893--934, 2019.

\bibitem{ahmed2013blind}
A.~Ahmed, B.~Recht, and J.~Romberg, ``Blind deconvolution using convex programming,'' \emph{IEEE Transactions on Information Theory}, vol.~60, no.~3, pp. 1711--1732, 2013.

\bibitem{kuo2019geometry}
H.-W. Kuo, Y.~Lau, Y.~Zhang, and J.~Wright, ``Geometry and symmetry in short-and-sparse deconvolution,'' in \emph{International Conference on Machine Learning}, 2019, pp. 3570--3580.

\bibitem{ahmed2018leveraging}
A.~Ahmed and L.~Demanet, ``Leveraging diversity and sparsity in blind deconvolution,'' \emph{IEEE Transactions on Information Theory}, vol.~64, no.~6, pp. 3975--4000, 2018.

\bibitem{scs}
B.~O'Donoghue, E.~Chu, N.~Parikh, and S.~Boyd, ``{SCS}: Splitting conic solver,'' \url{https://github.com/cvxgrp/scs}, 2019.

\bibitem{clup6}
M.~Stojnic, ``Regularly random duality,'' \emph{arXiv preprint arXiv:1303.7295}, 2013.

\bibitem{zhang2009music}
Y.~Zhang and B.~P. Ng, ``{MUSIC}-like {DOA} estimation without estimating the number of sources,'' \emph{IEEE Transactions on Signal Processing}, vol.~58, no.~3, pp. 1668--1676, 2009.

\bibitem{roy1989esprit}
R.~Roy and T.~Kailath, ``{ESPRIT}-estimation of signal parameters via rotational invariance techniques,'' \emph{IEEE Transactions on Acoustics, Speech, and Signal Processing}, vol.~37, no.~7, pp. 984--995, 1989.

\bibitem{starck2010sparse}
J.-L. Starck, F.~Murtagh, and J.~M. Fadili, \emph{Sparse image and signal processing: {W}avelets, curvelets, morphological diversity}.\hskip 1em plus 0.5em minus 0.4em\relax Cambridge University Press, 2010.

\bibitem{patel2002analysis}
P.~Patel and J.~Holtzman, ``Analysis of a simple successive interference cancellation scheme in a {DS/CDMA} system,'' \emph{IEEE Journal on Selected Areas in Communications}, vol.~12, no.~5, pp. 796--807, 2002.

\end{thebibliography}

\end{document}